\documentclass[12pt]{article}
\def \be {\begin{equation}}
\def \ee {\end{equation}}
\def \bea {\begin{eqnarray}}
\def \eea {\end{eqnarray}}
\def \nn {\nonumber}

\def \C {{\bf C}}

\def \del {\partial}
\def \dels {\partial\kern-.5em / \kern.5em}
\def \As {{A\kern-.5em / \kern.5em}}
\def \Ds {D\kern-.7em / \kern.5em}

\def \fe {\Psi}
\def \cfe {{\bar\Psi}}
\def \bo {\Phi}

\def \fm {Q}
\def \bm {P}
\def \pfm {q}
\def \pbm {p}

\def \f {f}
\def \h {h}

\def \a {\alpha}

\def \g {\gamma}

\def \d {\delta}

\def \lam {\lambda}

\def \Om {\Omega}
\def \th {\theta}

\def \cO {{\cal O}}
\def \bast {\bar{\ast}}

\begin{document}

\begin{titlepage}

\begin{center}

\hfill hep-th/0206077\\

\vskip .5in

\textbf{\Large Perturbative Approach to \\ Higher Derivative
Theories with Fermions  }

\vskip .5in
{\large Tai-Chung Cheng, Pei-Ming Ho, Mao-Chuang Yeh}
\vskip 15pt

{\small
Department of Physics, \\
National Taiwan University, \\
Taipei 106, Taiwan, \\
R.O.C.}\\

\vskip .2in

\sffamily{pmho@phys.ntu.edu.tw}

\vspace{60pt}

\end{center}

\begin{abstract}

We extend the perturbative approach
developed in an earlier work
to deal with Lagrangians which have arbitrary
higher order time derivative terms
for both bosons and fermions.
This approach enables us to find an effective Lagrangian
with only first time derivatives order by order
in the coupling constant.
As in the pure bosonic case,
to the first order,
the quantized Hamiltonian is bounded from below
whenever the potential is.
We show in the example of a single complex fermion
that higher derivative interactions
result in an effective mass and change of vacuum
for the low energy modes.
The supersymmetric noncommutative Wess-Zumino model
is considered as another example.
We also comment on the higher derivative terms
in Witten's string field theory
and the effectiveness of level truncation.

\end{abstract}

%\pacs{PACS numbers: 11.25.-w, 11.25.Mj, 11.25.Sq}%]

\end{titlepage}

\setcounter{footnote}{0}

\section{Introduction}

It is generally assumed that
the equation of motion for a bosonic variable
is a second order differential equation,
and that for a fermionic variable is first order.
Problems arise immediately when modifications
by higher derivative terms are introduced.
For example, even in the classical regime,
acausal behavior or runaway solutions appear
for charged point particles due to
radiation reaction which is a 3rd derivative term.
In general, canonical formulation \cite{Ostro}
always leads to a spectrum which is unbounded from below
for higher derivative theories.
For quantum field theories,
higher derivatives often imply
nonrenormalizability or violation of unitarity.
However, there are counter-examples \cite{Chu},
and in fact higher derivatives can be used to
improve the behavior of quantum field theories
by regularizing the ultra-violet divergences \cite{UV}.
Moreover, apart from the technical difficulty,
there is no known physical reason why
nature should abhor higher derivative interactions.
We are justified to ignore higher derivatives
only in the low energy limit,
and it would be a great puzzle if
we will never need higher derivative terms
in formulating fundamental theories.
In fact, in string field theory
there are indeed infinite higher derivatives.
For instance, in Witten's bosonic open string field theory \cite{Witten},
any field $f$ appears in the interaction as \cite{KS}
\be \label{ft}
\tilde{f} \equiv e^{a^2\del_{\mu}\del^{\mu}}f,
\ee
where $a^2 = \ln(3\sqrt{3}/4)\a'$.
See \cite{CHY} for references to
higher derivative (and nonlocal) theories
and their applications.

In a previous work \cite{CHY},
we considered a perturbative approach to
higher derivative theories,
which is equivalent to the approach of \cite{EW}.
It deals with Lagrangians whose kinetic term
is the same as that for ordinary free fields,
and all higher derivative terms appear only
in the interaction terms.
In a perturbative expansion of the coupling constant,
higher derivative terms are replaced by
lower derivative terms.
An effective Lagrangian is obtained in the end
with only first derivatives,
% new
and its quantization is straightforward.
This formulation is closely related to
the approach of Yang and Feldman \cite{YF}
which directly deals with the equations of motion
at the quantum level.
Recently, it was shown that the perturbative treatment
of field theories on noncommutative spacetime
in the spirit of \cite{YF} is unitary \cite{DFP,RY}.

The physical motivation for the perturbative approach is the following.
To extend the range of validity of a physical model
to higher energies,
we might need to add new interaction terms with
higher derivatives.
However, no matter how small the coupling is,
this would imply a sudden increase
in the dimension of phase space according to
the canonical formulation.
Furthermore, the Hamiltonian always becomes unbounded from below.
It is thus natural to carry out a projection
back to the original phase space,
which is called the ``reduced phase space'',
so that the new interaction term
will not abruptly change the theory into
a completely new theory that we don't know how to handle.
We showed that the Hamiltonian on the reduced phase space
is bounded from below at least to the first order \cite{CHY}
(whenever the potential is).

% new
To demonstrate the physical meaning of the perturbative approach
more explicitly, we showed in the example of a coupled spring system
\cite{CHY} that the perturbative approach gives
the correct description of the normal mode with lower natural frequency
while ignoring the other normal mode with higher natural frequency.

The purpose of this paper is to extend
our previous work to include fermions,
and to examine the effects of higher derivative interactions
on fermionic fields.
Interesting results are obtained.
Due to the anticommutativity of fermions,
the effect of higher derivative terms is
more severely restricted than bosons
in the perturbative approach.

For the case of a single complex fermion
in $0+1$ dimension, we find that, to all orders,
the effect of arbitrary higher derivative interactions
is always equivalent to a change of vacuum
and an effective mass.

A natural question about higher derivative theories
is whether supersymmetry can cure the problem of
the Hamiltonian being unbounded from below.
We will show in an example that in the canonical formulation,
the quantization of fermions with higher derivatives
will destroy the reality conditions.
Thus the supercharge $Q$ will not be Hermitian,
and we can not conclude that $H \geq 0$ from
the SUSY algebra $H \sim Q^2$.
We will apply the perturbative approach to
the $1+1$ dimensional supersymmetric
noncommutative Wess-Zumino model \cite{WZ} as an example.
The perturbative approach not only provides a consistent
quantization but also preserves supersymmetry.

Another interesting observation we make in this paper is that
the higher derivatives in (\ref{ft})
effectively increase the interaction strength
for fields of higher levels.
This makes the technique of level truncation
less effective for fluctuation modes
compared with the zero modes in the string field theory.

This paper is organized as follows.
We show in sec.\ref{Canonical} that
higher derivative terms introduce a different problem
for fermions in the canonical formulation.
The problem is that canonical quantization is not
consistent with the reality (Hermiticity) condition on fields.
In sec.\ref{Perturb},
we extend our previous work to the generic case
with arbitrary numbers of bosons and fermions.
Explicit results to the first order are given,
showing that the Hamiltonian becomes bounded from below
when the potential is.
We describe how to carry out
this procedure to an arbitrary order.
We prove that the perturbative approach
is consistent to all orders and that
the effect of higher derivative terms
is properly preserved in this approach.
The case of a single complex fermion in 0+1 dimension
is considered as a simple example.
We find that the effect of any higher derivative interactions
for this case sums up to
a change of vacuum
and an effective mass, to arbitrary orders.
We also consider the noncommutative Wess-Zumino model
to the first order as another example.
For Witten's open string field theory,
we find that the higher derivatives in (\ref{ft})
make level truncation less effective for fluctuation modes.
These examples are given in sec.\ref{Examples:}.
Finally, in sec.\ref{Extension:},
we further extend our approach to
a larger reduced phase space which contains
fields with derivatives up to any finite order.

\section{Canonical Formulation} \label{Canonical}

Consider the canonical formulation with higher time derivatives
for both bosons and fermions. The Lagrangian is
\begin{equation}
L_0(\bo_a^{(i)},\fe_{a'}^{(j)}),
\end{equation}
where $\bo^{(i)}_a$ $(\fe^{(j)}_{a'})$ is the $i(j)$-th time
derivative of the $a$$(a')$-th boson (fermion),
$a(a')=1\cdots M_B (M_F)$ and $i(j)=1\cdots N_B (N_F)$.

Let us apply the formalism of \cite{FJ}. The variation of the
action
\[
S=\int_{t_i}^{t_f} dt \; L_0
\]
with respect to $\bo_a$,$\fe_{a'}$ is found to be the time
integral of the Euler-Lagrange equations
\be \label{EOM0}
\sum_{i=0}^{N_B} \left(-\frac{d}{dt}\right)^i
\frac{\del L_0}{\del \bo_a^{(i)}} = 0,
\ee
\be
\sum_{j=0}^{N_F} \left(-\frac{d}{dt}\right)^j \frac{\del L_0}{\del
\fe_{a'}^{(j)}}=0 \ee multiplied by $\d \bo_a$, $\d \fe_{a'}$ from
the right, plus a boundary term
\be
\left[ \sum_{a=1}^{M_B}\sum_{i=0}^{N_B-1} \bm_{ai} \d \bo^{(i)}_a
+\sum_{a'=1}^{M_F}\sum_{j=0}^{N_F-1} \fm_{a'j} \d \fe_{a'}^{(j)}
\right]_{t_i}^{t_f},
\label{bun}
\ee
where $\bm_{ai}$ and $\fm_{a'j}$
are the conjugate momenta of $\bo^{(i)}_a$ and $\fe^{(j)}_{a'}$
given by
\begin{eqnarray}
\bm_{ai}&=& \sum_{k=0}^{N_B-i-1}\left(-\frac{d}{dt}\right)^k
\frac{\del L_0}{\del \bo _a^{(i+k+1)}}, \label{bm1}
\\
\fm_{a'j}&=&\sum_{h=0}^{N_F-j-1}\left(-\frac{d}{dt}\right)^h
\frac{\del L_0}{\del \fe_{a'}^{(j+h+1)}}. \label{fm1}
\end{eqnarray}
In the above, $\frac{\partial}{\partial \fe}$ is defined to be the
derivative with respect to $\Psi$ from the right. The symplectic
structure can be directly read off from the boundary term
(\ref{bun}) as
\be
\Om=\sum_{a=1}^{M_B}\sum_{i=0}^{N_B-1} d\bm_{ai} d
\bo_a^{(i)}+\sum_{a'=1}^{M_F}\sum_{j=0}^{N_F-1} d\fm_{a'j} d
\fe_{a'}^{(j)}.
\label{sym2}
\ee
Note that $d\Phi$'s are
anticommuting variables while $d\Psi$'s are commuting variables.

In the canonical formulation, the Hamiltonian is
\be
H=\sum_{a=1}^{M_B}\sum_{i=0}^{N_B-1} \bm_{ai}
\bo_{a}^{(i)}+\sum_{a'=1}^{M_F}\sum_{j=0}^{N_F-1} \fm_{a'j}
\fe_{a'}^{(j)} - L_0.
\ee
Assuming that the action is nondegenerate,
that is, the definition of $P_{a(N_B-1)}$ in (\ref{bm1})
can be used to solve $\Phi_a^{N_B}$ as a function of
$P_{a(N_B-1)}$ and $\Phi$'s.
This implies that the Hamiltonian is unbounded from below
because it is linear in all $P_{ai}$ for $i<(N_B-1)$.
Thus we expect violation of unitarity after standard
quantization of the system.

Now we consider the fermions.
The classical Hamiltonian of a fermion
is typically unbounded from below even
without higher derivatives. Yet canonical quantization leads to a
spectrum which is bounded from below by filling the Dirac sea.
This difference between the bosons and fermions raises a puzzle.
What happens if we have a higher derivative theory with
supersymmetry, where the bosonic spectrum should be identical to
the fermionic spectrum?

It turns out that the problem of femions with higher time
derivatives is that the canonical quantization is inconsistent
with the reality conditions, and thus violation of unitarity is
also expected. Let us consider the simplest example of a real
fermion $\psi$ in 0+1 dimension with the Lagrangian
\be
L_0=i\psi\dot{\psi}+ig\dot{\psi}\ddot{\psi}.
\ee
The sympletic form is
\be
\Om=i\left(d\phi d\phi+gd\dot{\psi}d\dot{\psi}
-g^2 d\ddot{\psi}d\ddot{\psi}\right),
\ee
where $\phi\equiv \psi-g\ddot{\psi}$.
The Poisson brackets are
\be
(\phi,\phi)=i, \quad (\dot{\psi},\dot{\psi})=i/g,
\quad (\ddot{\psi},\ddot{\psi})=-i/g^2.
\ee
Upon quantization, we replace Poisson brackets by
anticommutators for fermions, up to a factor of $\pm i$.
For the two possible choices of sign
\be
\{\cdot, \cdot\} = i(\cdot, \cdot), \quad \mbox{or} \quad
\{\cdot, \cdot\} = -i(\cdot, \cdot),
\ee
we will have either
\be
\phi^2 < 0, \quad \mbox{or} \quad \ddot{\psi}^2 < 0,
\ee
which is
inconsistent with the fact that $\psi$ is real.
This implies that the supercharge $Q$ is not Hermitian.
In fact the quantization is simply inconsistent.
Hence we find that the problem for fermions with
higher derivative Lagrangians can be even more serious than bosons.

\section{Perturbative Approach} \label{Perturb}

In this paper we focus on Lagrangians of the form
\be
L=L_B+L_F-\lambda V,
\ee
where
$\lam$ is the coupling constant,
$L_B$ and $L_F$ are the free field Lagrangians
for bosons and fermions,
and $V$ is the interaction piece
where higher derivative terms reside.
More explicitly, we have
\begin{eqnarray}
L_B&=&\sum_a
\frac{1}{2}\dot{\bo}_a^2-\sum_{a}\frac{1}{2}m^{B}_{ab} \bo_a\bo_b,
\\
L_F&=&\sum_{a'}
i\frac{1}{2}\fe_{a'}\dot{\fe}_{a'}-\sum_{a',b'}
i\frac{1}{2}m^{F}_{a'b'}\fe_{a'}\fe_{b'},
\\
V&=&V(\bo_a^{(i)},\fe_{a'}^{(j)}),
\end{eqnarray}
where $m^B$ is symmetric and $m^F$ is antisymmetric.
We will give the prescription for the perturbative approach
for such Lagrangians in this section.

To be general, let us consider the cases with infinite order time
derivatives.
Under variation the action is
\be
\delta S=-\int dt
\left(\sum_a(EOM_B)_a\delta\bo_a+\sum_{b'}(EOM_F)_{b'}
\delta\fe_{b'}\right)+\left[\sum_{k=0}^{\infty}\bm_{ak}
\delta\bo_a^{(k)}+\sum_{h=0}^{\infty}\fm_{b'h}
\delta\fe_{b'}^{(h)}\right]^{t_f}_{t_i},
\ee
where the equations
of motion for bosons and fermions are
\begin{eqnarray}
(EOM_B)_a&\equiv&\ddot{\bo}_a+\sum_{b}m^B_{ab}\bo_b
+\lambda\sum_{i=0}^{\infty}\left(-\frac{d}{dt}\right)^i
\frac{\partial V}{\partial \bo_a^{(i)}}, \label{EOMb}
\\
(EOM_F)_{a'}&\equiv&i\dot{\fe}_{a'}- i\sum_{b'}m^F_{a'b'}\fe_{b'}
+\lambda\sum_{i=0}^{\infty}\left(-\frac{d}{dt}\right)^i
\frac{\partial V}{\partial \fe_{a'}^{(i)}} \label{EOMf}.
\end{eqnarray}
According to (\ref{bm1}) and (\ref{fm1}),
the canonical momenta $\bm_a$ and $\fm_{b'}$ are
\begin{eqnarray}
\bm_{ak}&=&\dot{\bo}_a\delta_{k0}-\lambda
\sum_{i=k+1}^{\infty}\left(-\frac{d}{dt}\right)^{(i-k-1)}
\frac{\partial V}{\partial \bo_a^{(i)}}, \label{bm}
\\
\fm_{a'k}&=&\frac{1}{2}{\fe_{a'}}\delta_{k0}-\lambda
\sum_{i=k+1}^{\infty}\left(-\frac{d}{dt}\right)^{(i-k-1)}
\frac{\partial V}{\partial \fe_{a'}^{(i)}}. \label{fm}
\end{eqnarray}
The symplectic two-form is given by (\ref{sym2}).
%\be
%\Omega=\sum^{\infty}_{k=0}\left(\sum_a
%d\bm_{ak}d\bo_a^{(k)}+\sum_{b'}d\fm_{b'k}d\fe^{(k)}_{b'}\right)
%\label{omega} \ee

\subsection{First Order Approximation}

Following \cite{CHY},
we will construct the effective action without higher derivatives
for a reduced phase space which is appropriate for
the low energy, weak coupling regime.
We will keep $\fe$ for fermions and $\bo$,
$\dot{\bo}$ for bosons as the variables for our reduced phase space.
Our strategy is to use the equations of motion to replace higher
derivative terms by lower derivative ones.
To the lowest order,
\begin{eqnarray}
&\ \bo^{(n)}_a&\simeq \left\{\begin{array}{ll}
  \sum_bM^{B(\frac{n}{2})}_{ab}\bo_b & (n=\mbox{even}), \nn \\
  \sum_bM^{B(\frac{n-1}{2})}_{ab}\dot{\bo}_b & (n=\mbox{odd}), \nn\\
  \end{array}\right.
\\
&\fe^{(n)}_{a'}&\simeq\sum_{b'}M_{a'b'}^{F(n)}\fe_{b'},
\label{iterf}
\end{eqnarray}
where
\be
M^{B(n)}_{a_na_0}\equiv \left\{\begin{array}{ll}
         \sum_{a_{n-1}}\cdots\sum_{a_{1}}(-m^{B}_{a_{n}a_{n-1}})\cdots
             (-m^{B}_{a_{1}a_{0}})  & (n\geq 2), \\
         -m^{B}_{a_1a_0} & (n=1),\\
         \delta_{a_n a_0}&(n=0);
       \end{array}\right.
\label{deltb}
\ee
\be
M^{F(n)}_{a'_na'_0}\equiv \left\{\begin{array}{ll}
         \sum_{a'_{n-1}}\cdots\sum_{a'_{1}}(m^F_{a'_{n}a'_{n-1}})\cdots
         (m^F_{a'_{1}a'_{0}})  & (n\geq 2), \\
         m_{a'_1a'_0} & (n=1),\\
         \delta_{a'_n a'_0}&(n=0).
       \end{array}\right.
\label{deltf}
\ee
The symplectic two-form (\ref{sym2}) reduces to
\be
[\Omega]_1=\sum_a d\pbm_{a0}d \bo_a+d\pbm_{a1}d\dot{\bo}_a
+\sum_{a'}d\pfm_{a'}d\fe_{a'},
\ee
where
\be
\pbm_{a0} = \dot{\bo}_a-\lambda\xi_{a0}, \quad \pbm_{a1} =
-\lambda\xi_{a1}, \quad \pfm_{a'} = \frac{i}{2}\fe_{a'}-\lambda
\xi_{a'2},
\ee
and
\bea
\xi_{a0}&=&\sum_{b=1}^{M_B}\sum_{j=0}^{\infty}\sum_{i=2j+1}^{\infty}
M_{ab}^{B(j)}\left[\left(-\frac{d}{dt}\right)^{i-2j-1} \frac{\del
V}{\del \bo_b^{(i)}}\right]_1,
\label{xi0}
\\
\xi_{a1}&=&\sum_{b=1}^{M_B}\sum_{j=0}^{\infty}\sum_{i=2j+2}^{\infty}
M_{ab}^{B(j)}\left[\left(-\frac{d}{dt}\right)^{i-2j-2} \frac{\del
V}{\del \bo_b^{(i)}}\right]_1, \label{xi1} \\
\xi_{a'2}&=&\sum_{b'=1}^{M_F}\sum_{j=0}^{\infty}\sum_{i=j+1}^{\infty}
M_{b'a'}^{F(j)}\left[\left(-\frac{d}{dt}\right)^{i-j-1}
\frac{\del V}{\del \fe_{b'}^{(i)}}\right]_1 .
\label{xi2}
\eea
Here $[\cdot]_1$ refers to the replacement
of higher derivative terms by functions of
$\bo$, $\dot{\bo}$ and $\dot{\fe}$ via (\ref{iterf}).

Explicitly, the symplectic two-form is
\begin{eqnarray}
[\Omega]_1 &=& \sum_{ab}\left\{ \left[ -\delta_{ab}+\lambda \left(
\frac{\partial\xi_{a0}}{\partial\dot{\bo}_{b}}-
\frac{\partial\xi_{b1}}{\partial\bo_a}\right) \right] d\bo_a
d\dot{\bo}_b+ \lambda \frac{\partial\xi_{a0}}{\partial\bo_b}d\bo_a
d\bo_b +\lambda\frac{\partial\xi_{a1}}{\partial\dot{\bo}_b}
d\dot{\bo}_a d\dot{\bo}_b \right\} \nn \\ &+& \sum_{a'b'}
\left[i\frac{\delta_{a'b'}}{2}-
\lambda\frac{\partial_R\xi_{a'2}}{\partial\fe_{b'}}\right]
d\fe_{a'}d\fe_{b'} \nn \\ &+&\sum_{a,a'}\left\{
\lambda\left(\frac{\partial_R\xi_{a0}}{\partial\fe_{a'}}-
\frac{\partial\xi_{a'2}}{\partial\bo_a}\right)d\bo_ad\fe_{a'}
+\lambda \left(\frac{\partial_R\xi_{a1}}{\partial\fe_{a'}}-
\frac{\partial\xi_{a'2}}{\partial\dot{\bo}_a}\right)d\dot{\bo}_a
d\fe_{a'}\right\}.
\end{eqnarray}
Inverting the symplectic 2-form,
we find the Poisson brackets
to the lowest order in $\lam$
\begin{eqnarray}
(\bo_a,\dot{\bo}_b)&=& \delta_{ab} +\lambda \left(
\frac{\partial\xi_{b0}}{\partial\dot{\bo}_a}
-\frac{\partial\xi_{a1}}{\partial\bo_b}\right), \\
(\bo_a,\bo_b)&=&\lambda
\left(\frac{\partial\xi_{a1}}{\partial\dot{\bo}_b}
-\frac{\partial\xi_{b1}}{\partial\dot{\bo}_a}\right), \\
(\dot{\bo}_a,\dot{\bo}_b) &=&
\lambda\left(\frac{\partial\xi_{a0}}{\partial\bo_b}-
\frac{\partial\xi_{b0}}{\partial\bo_a}\right), \\
(\fe_{a'},\fe_{b'})&=&-i\delta_{a'b'}-
\lambda\left(\frac{\partial\xi_{b'2}}{\partial\fe_{a'}}+
\frac{\partial\xi_{a'2}}{\partial\fe_{b'}}\right), \\
(\bo_a,\fe_{b'}) &=&
-i\lambda\left(\frac{\partial\xi_{b'2}}{\partial\dot{\bo}_a}
-\frac{\partial\xi_{a1}}{\partial\fe_{b'}}\right), \\
(\dot{\bo}_a,\fe_{b'}) &=& i\lambda\left(
\frac{\partial\xi_{b'2}}{\partial\bo_a}
-\frac{\partial\xi_{a0}}{\partial\fe_{b'}}\right).
\end{eqnarray}
Remarkably, by a simple change of variables
\be
\label{mom}
\varphi_a=\bo_a+\lambda \xi_{a1},\quad
\pi_a=\dot{\bo}_a-\lambda \xi_{a0}, \quad
\psi_{a'}=\fe_{a'}+i\lambda \xi_{a'2},\ee
the Poisson brackets can be put
in the standard form
\begin{eqnarray}
(\varphi_a,\pi_b)=\delta_{ab}, \quad
(\psi_{a'},\psi_{b'})=-i\delta_{a'b'},
\end{eqnarray}
with all other Poisson brackets vanishing.

The Hamiltonian for the reduced phase space variables
is defined as
\be
\tilde{H}_1=\left[\sum_a(p_{a0}\dot{\bo}_a+p_{a1}\ddot{\bo}_a)
+\sum_{a'}q_{a'}\dot{\fe}_{a'} - L\right]_1.
\label{H1}
\ee
In terms of the new variables $\varphi$,
$\pi$ and $\psi$, it is
\be
\tilde{H}_1=\sum_a \frac{1}{2}\pi_a^2+\sum_{a,b}\frac{1}{2}m^{B}_{ab}
\varphi_a\varphi_b+ \sum_{a',b'}
\frac{i}{2}m^{F}_{a'b'}\psi_{a'}\psi_{b'} +\lambda
[V]_1(\varphi,\pi,\psi).
\ee

Note that, if the potential V is bounded from below,
the first order Hamiltonian is also bounded from below.
One can check that the Hamilton equations
\be
\dot{\varphi}_a=(\varphi_a,\tilde{H}_1), \quad
\dot{\pi}_a=(\pi_a,\tilde{H}_1),\quad
\dot{\psi}_{a'}=(\psi_{a'},\tilde{H}_1)
\label{E1}
\ee
reproduce the equations of motion (\ref{EOMb}), (\ref{EOMf})
to the first order in $\lambda$.

For the reduced phase space,
the effective Lagrangian
corresponding to the Hamiltonian (\ref{E1}) is found to be
\be
\label{effL}
\tilde{L}_1=\frac{1}{2}\sum_a\dot{\varphi}_a^2-
\sum_{a,b}\frac{m^B_{ab}}{2}\varphi_a\varphi_b+\sum_{a'}
\frac{1}{2}i\psi_{a'}\dot{\psi}_{a'}-\sum_{a'b'}
\frac{1}{2}im^{F}_{a'b'}\psi_{a'}\psi_{b'}-\lambda
[V]_1(\varphi,\dot{\varphi},\psi).
\ee
Its Euler-Lagrange equation
agrees with the original system to the first order in $\lambda$.
For this construction of the effective Lagrangian to be
self-consistent, we also need the conjugate momenta of $\varphi$
and $\psi$ defined from the effective Lagrangian (\ref{effL}) to
agree with (\ref{mom}). While the consistency for the fermions is
trivial, for the bosons we need to use the identity
\be
% new
\xi_{a0}+[\dot{\xi}_{a1}]_1 = \frac{\del [V]_1}{\del \dot{\Phi}_a},
\ee
which can be verified using (\ref{xi0}) and (\ref{xi1}). Since the
final expression of the Lagrangian (\ref{effL}) contains only
first derivatives, its quantization is straightforward.

The general result (\ref{effL}) is very useful.
It says that, to the first order approximation,
in terms of some new variables,
the effective Lagrangian is formally the same
as simply reducing all higher dervative terms
in the original Lagrangian to lower derivatives
according to the free field equations.

As Lagrangians are only defined up to total derivatives,
let us comment on the difference total derivative terms can make.
Starting with two Lagrangians differing from
each other only by total derivatives,
their effective actions (\ref{effL}) will appear to be different.
But this difference simply originates from a different definition
of the variables $\varphi, \psi$,
and the effective Lagrangians are in fact equivalent.

\subsection{Higher Order Approximation}

For higher order corrections,
we first iterate the equations of motion
(\ref{EOMb}), (\ref{EOMf}) up to a certain order $\cO(g^n)$.
For example, to the first order,
\begin{eqnarray}
\ddot{\bo}_a&\rightarrow& -\sum_bm^B_{ab}
\bo_b-\lambda\sum_{i=0}^{\infty} \left[
\left(-\frac{d}{dt}\right)^i\frac{\del V}{\del
\bo_a^{(i)}} \right]_1,
\label{q1}
\\
\dot{\fe}_{a'}&\rightarrow& \sum_{b'}
m^F_{a'b'} \fe_{b'}+i\lambda\sum_{j=0}^{\infty} \left[
\left(-\frac{d}{dt}\right)^j\frac{\del V}{\del
\fe_{a'}^{(j)}} \right]_1.
\label{q2}
\end{eqnarray}
%where we should also use the replacement (\ref{iterf}) to change
%the last terms in both lines into a function of $\bo_a$,
%$\dot{\bo}_a$ and $\fe_{a'}$ only.
Higher derivatives of $\bo_a$,
$\fe_{a'}$ can also be replaced by functions of $\bo_a$,
$\dot{\bo}_a$ and $\fe_{a'}$ up to the same order in $\lambda$ by
differentiating with respect to time and repeatedly using
(\ref{q1}), (\ref{q2}) as
\begin{eqnarray}
\bo_a^{(n)}&\simeq&\left\{
\begin{array}{ll}
\sum_b\left( M^{B(n/2)}_{ab}\bo_b
-\lambda\sum_{l=1}^{n/2}\sum_{k=0}^{\infty} \left[
M^{B(n/2-l)}_{ab}\left(-\frac{d}{dt}\right)^{k+2l-2} \frac{\del
V}{\del \bo_b^{(k)}} \right]_1 \right) & (n=\mbox{even}),
\\ \sum_b \left(
M^{B(n-1)/2}_{a,b}\dot{\bo}
+\lambda\sum_{l=1}^{(n-1)/2}\sum_{k=0}^{\infty} \left[
M^{B((n-1)/2-l)}_{ab}\left(-\frac{d}{dt}\right)^{k+2l-1}
\frac{\del V}{\del \bo_b^{(k)}} \right]_1 \right) &
(n=\mbox{odd}),\\
\end{array}
\right.
\end{eqnarray}
\begin{eqnarray}
\fe_{a'}^{(n)}&\simeq& \sum_{b'}\left( M^{(n)}_{a'b'}\fe_{b'}
+i\lambda\sum_{j=0}^{\infty}\sum_{l=1}^{n} \left[ M_{a'b'}^{F(n-l)}
\left(-\frac{d}{dt}\right)^{j+l-1} \frac{\partial V} {\partial
\fe_{b'}^{(j)}} \right]_1 \right).
\end{eqnarray}
In general, we can always have all
$\bo_a^{(n)}$, $\fe_{a'}^{(n)}$ expressed as functions of $\bo_a$,
$\dot{\bo}_a$ and $\fe_{a'}$ only, up to any given order ${\cal
O}(\lambda^p)$.
This helps us to derive the effective symplectic form
from (\ref{sym2})
%\bea
\be
[\Omega]_p
%&=&
=\sum_{a,a'}
\left[d\bm_{ai}d\bo_a^{(i)}+d\fm_{a'i}d\fe_{a'}^{(i)}\right]_p ,
%\nn
%\\ &=&\sum_{a,a'}\left[d\pbm_{0a}d\bo_a+d\pbm_{1a}\dot{\bo}_a
%+d\pfm_{a'}d\fe_{a'}\right] \nn \\ &=&\sum_{a',b}\frac{1}{2}\left(
%\frac{\del\pbm_{0b}}{\del\bo_a}
%-\frac{\del\pbm_{0a}}{\del\bo_b}\right)d\bo_ad\bo_b+
%\left(\frac{\del\pbm_{1b}}{\del\bo_a}
%-\frac{\del\pbm_{0a}}{\del\dot{\bo}_b}\right)d\bo_ad\dot{\bo}_b
%\nn \\ &+& \left(\frac{\del\pfm_{b'}}{\del\bo_a}-
%\frac{\del\pbm_{0a}}{\del\fe_{b'}}\right)d\bo_ad\fe_{b'}+
%\left(\frac{\del\pfm_{b'}}{\del\dot{\bo}_a}
%-\frac{\del\pbm_{1a}}{\del\fe_{b'}}\right)d\dot{\bo}_ad\fe_{b'}
%\nn \\ &+& \frac{1}{2}\left(\frac{\del\pbm_{1b}}{\del\dot{\bo}_a}
%-\frac{\del\pbm_{1a}}{\del\dot{\bo}_b}\right) d\dot{\bo}_a
%d\dot{\bo}_b+ \frac{1}{2}\left(\frac{\del\pfm_{b'}}{\del\fe_{a'}}
%+\frac{\del\pfm_{a'}}{\del\fe_{b'}}\right)d\fe_{a'}d\fe_{b'},
\label{sym1}
%\eea
\ee
where the bracket $[\;\cdot\;]_p$ means to replace
all higher derivatives of $\bo_a$ and $\fe_b$ by
functions of $\bo_a$, $\dot{\bo}_a$ and $\fe_b$
up to order $\lam^p$.
%using the equations of motion as we described above
%to the order $\lam^p$.
The final Hamiltonian is defined by (\ref{H1})
with $[\;\cdot\;]_1$ replaced by $[\;\cdot\;]_p$.
The Hamilton equations will give the equations of motion up to
$\cO(\lambda^{p+1})$.

\subsection{To All Orders: A Formal Proof} \label{AllOrder}

Now we give a formal proof for the self-consistency of the
perturbative formulation. {}From the equations of motion
(\ref{EOMb}), (\ref{EOMf}), assume that one can find
an exact solution (to all orders in $\lambda$)
\be
\label{qf}
\ddot{\bo}_a=\h_a(\bo_b,\dot{\bo}_b,\fe_{b'}), \quad
\dot{\fe}_{a'}=\f_{a'}(\bo_b,\dot{\bo}_b,\fe_{b'})
\ee
for certain functions $\f$ and $h$
by infinite iteration (or inspiration).
{}From this, higher derivatives of $\bo_a
$ and $\fe_{a'}$ can be written as functions
on the reduced phase space
\be
\label{hfn}
\bo_a^{(i)}=\h_{ai}(\bo_b,\dot{\bo}_b,\fe_{b'}), \quad
\fe_{a'}^{(i)}=\f_{a'i}(\bo_b,\dot{\bo}_b,\fe_{b'}).
\ee
The functions $\h_{ai}$ and $\f_{a'i}$
can be obtained recursively
\be
\h_{a(i+1)}=\left[\frac{d}{dt}\h_{ai}\right]=\left( \frac{\del
\h_{ai}}{\del \bo_b}\dot{\bo}_b +\frac{\del \h_{ai}}{\del
\dot{\bo}_b}\h_b +\frac{\del \h_{ai}}{\del
\fe_{b'}}\f_{b'}\right),
\ee
\be
\f_{a'(i+1)}=\left[\frac{d}{dt}\f_{a'i}\right]= \left(\frac{\del
\f_{a'i}}{\del \bo_b} \dot{\bo}_b+\frac{\del \f_{a'i}}{\del
\dot{\bo}_b}\h_b +\frac{\del \f_{a'i}}{\del
\fe_{b'}}\f_{b'}\right),
\ee
where we used the Einstein's
summation convension and the notation
\be
[A]\equiv A|_{%\left
\{\bo_a^{(i)}=\h_{ai} \ , \ %\quad
\fe_{a'}^{(i)}=\f_{a'i}%\right
\} }. \ee

A few identities that will come in handy in the proof
are the following. From (\ref{bm1}), (\ref{fm1}) we find
\be
\dot{\bm}_{ai}=\frac{\del L_0}{\del \bo_{a}^{(i)}}-\bm_{a
(i-1)},\quad \dot{\fm}_{a'i}=\frac{\del L_0}{\del
\fe_{a'}^{(i)}}-\fm_{a'(i-1)}.
\label{rel1}
\ee
For an arbitrary function $A$ on the total phase space,
we have the following identities
\bea
\frac{d}{dt}[A]&=&\left[\dot{A}\right]+ \frac{\del
[A]}{\del\dot{\bo}_a}(\ddot{\bo}_a-\h_a) + \frac{\del
[A]}{\del\fe_{a'}} (\dot{\fe}_{a'}-\f_{a'}), \label{rel2}
\\
\frac{\del [A]}{\del \fe_{a'}}&=&\left[\frac{\del A}{\del
\bo^{(i)}_b }\right] \frac{\del \h_{bi}}{\del
\fe_{a'}}+\left[\frac{\del A}{\del \fe^{(i)}_{b'} }\right]
\frac{\del f_{{b'}i}}{\del \fe_{a'}}, \label{rel3} \eea where one
can also replace $\del/\del\fe_{a'}$ \ by $\del/\del\bo_{a}$ \ or
\ $\del/\del\dot{\bo}_{a}$ in the last formula.
The effective conjugate momenta
$\pbm_{0a}$, $\pbm_{1a}$, $\pfm_{a'}$ are defined by
\be
\left[\sum_{a,a',i}\bm_{ai}\d\bo_a^{(i)}
+\fm_{a'i}\d\fe_{a'}^{(i)}\right]=
\sum_{a,a'}\pbm_{0a}\d\bo_a
+\pbm_{1a}\d\dot{\bo}_a+\pfm_{a'}\d\fe_{a'},
\ee
and they are
\be
\pbm_{0a}=\left[ \bm_{bj}\frac{\partial h_{bj}}{\partial\bo_a}
+\fm_{b'j}\frac{\partial f_{b'j}}{\partial\bo_a}\right],
\label{q0a}
\ee
\be
\pbm_{1a}=\left[ \bm_{bj}\frac{\partial
h_{bj}}{\partial\dot{\bo}_a} +\fm_{b'j}\frac{\partial f_{b'j}
}{\partial\dot{\bo}_a}\right],
\label{q1a}
\ee
\be
\pfm_{a'}=\left[ \bm_{bj}\frac{\partial h_{bj}}{\partial\fe_{a'}}
+\fm_{b'j}\frac{\partial f_{b'j}}{\partial\fe_{a'}}\right].
\label{pa}
\ee
%\be
%\pfm_{0,a} = \left[
%\bm_{b,j}\frac{\partial\bo^{(j)}_b}{\partial\bo_a}
%+\fm_{b,j}\frac{\partial\fe^{(j)}_b}{\partial\fe_a}\right]
%\ee

The effective Hamiltonian is
\be \label{eH}
H=\left[\pbm_{0a}\dot{\bo}_a+\pbm_{1a}\ddot{\bo}_a
+\pfm_{a'}\dot{\fe}_{a'}-L \right].
\ee
The Hamilton equations
%(\ref{E1})
based on the symplectic structure (\ref{sym1}) are
\bea
(\pbm_{0a})^.&=& -\left(\pbm_{1b}\frac{\del\h_b}{\del\bo_a}
+\pfm_{b'}\frac{\del\f_{b'}}{\del\bo_a}\right) +\frac{\del
[L]}{\del\bo_a}
+\frac{\del\pbm_{1b}}{\del\bo_a}(\ddot{\bo}_b-\h_b)
+\frac{\del\pfm_{b'}}{\del\bo_a}(\dot{\fe}_{b'}-\f_{b'}),
\label{heq0}
\\
(\pbm_{1a})^.&=&
-\left(\pbm_{1b}\frac{\del\h_b}{\del\dot{\bo}_a}
+\pfm_{b'}\frac{\del\f_{b'}}{\del\dot{\bo}_a}\right)
+\frac{\del [L]}{\del\dot{\bo}_a}
+\frac{\del\pbm_{1b}}{\del\dot{\bo}_a}(\ddot{\bo}_b-\h_b)
+\frac{\del\pfm_{b'}}{\del\dot{\bo}_a}(\dot{\fe}_{b'}-\f_{b'})
-\pbm_{0a},
\label{heq1}
\\
(\pfm_{a'})^.&=& -\left(\pbm_{1b}\frac{\del\h_b}{\del\fe_{a'}}
+\pfm_{b'}\frac{\del\f_{b'}}{\del\fe_{a'}}\right) +\frac{\del
[L]}{\del\fe_{a'}}
+\frac{\del\pbm_{1b}}{\del\fe_{a'}}(\ddot{\bo}_b-\h_b)
-\frac{\del\pfm_{b'}}{\del\fe_{a'}}(\dot{\fe}_{b'}-\f_{b'}).
\label{hepa}
\eea
With the help of (\ref{rel1})-(\ref{rel3}), one
can show from (\ref{q0a}), (\ref{q1a}) and (\ref{pa}) that
(\ref{heq0}) is automatically satisfied, and that (\ref{heq1}),
(\ref{hepa}) are equivalent to the equations of motion (\ref{qf}).

\section{Examples} \label{Examples:}

\subsection{A Single Complex Fermion}

Let's consider
the case of a single complex fermion in $0+1$ dimension
as a simple example.
Remarkably, in this case we can describe the effect
of arbitrary higher derivative interactions to all orders.
%\be
%\cfe=\fe^{*} \ \Rightarrow \
%L(\cfe^{(i)},\fe^{(i)})
%\ee
%As in the previous discussion, we find
%where
%\be
%\cfm_i=\Sigma^n_{j=i+1}(-\frac{d}{dt})^{j-i-1} \frac{\delta_L
%L}{\delta \cfe^{(j)} } , \quad
%\fm_i=\Sigma^n_{j=i+1}(-\frac{d}{dt})^{j-i-1} \frac{\delta_R
%L}{\delta \fe^{(j)} } . \ee
Assume that the Lagrangian is of the following form
\be
L=\frac{i}{2}(\cfe\dot{\fe}+\fe\dot{\cfe})
-m\cfe \fe -\lambda V(\cfe^{(i)},\fe^{(i)}),
\ee
where $\cfe$ is the complex conjugation of $\fe$.
Instead of decomposing the complex fermion into two real fermions,
we will maintain the complex structure.
By applying integration by parts to the action,
we can always rewrite the Lagrangian in such a way that
$\fe \leftrightarrow \cfe$ is a symmetry.

%The variation of the action is
%\be
%\delta S=\left[\delta\cfe^{(i)}\cfm_i +\fm_i\delta
%\fe^{(i)}\right]_{t_i}^{t_f} +
%\int dt \left( \d\cfe(EOM) + (\overline{EOM}) \d\fe \right).
%\left\{\delta\cfe(\Sigma^n_{i=0}(-\frac{d}{dt})^i \frac{\delta_L
%L }{\delta \cfe^{(i)}})+ (\Sigma^n_{i=0}(-\frac{d}{dt})^i
%\frac{\delta_R L }{\delta \fe^{(i)}}) \delta \fe \right\}.
%\label{dS01}
%\ee
%The conjugate momenta are
%\be
%\cfm_i=-\lambda\sum^n_{j=i+1}
%(-\frac{d}{dt})^{j-i-1}\frac{\partial_L V}{\partial \cfe^{(j)} } ,
%\quad \fm_i=i\cfe\delta_{i0}-\lambda\sum^n_{j=i+1}
%(-\frac{d}{dt})^{j-i-1} \frac{\partial_R V}{\partial \fe^{(j)} } ,
%\ee
%where

%Derivatives of $\fe$ will be taken from the right,
%and those of $\cfe$ from the left.

The equations of motion are
\be
i\dot{\fe}-m\fe-\lambda\sum^n_{i=0}
\left(-\frac{d}{dt}\right)^i\frac{\partial_L V}{\partial \cfe^{(i)}} =0,
\label{EOM01}
\ee
and its complex conjugation.
The subscript $L$ of $\del_L$ refers to differentiation from the left.

%\be
%-i\dot{\cfe}-m\cfe-\lambda\sum^n_{i=0} (-\frac{d}{dt})^i
%\frac{\partial_R V}{\partial \fe^{(i)}}=0 .
%\label{EOM01c}
%\ee

%\subsection{To All Orders}

It is straightforward to see that,
by iterating the equations of motion order by order in $\lam$,
the function $f$ defined in (\ref{qf})
will be of the following form
\begin{eqnarray}
\left (\begin{array}{ll} \dot{\fe} \\ \dot{\cfe}
\end{array}\right)
& = \left (\begin{array}{ll} R_1&R_2 \\ \bar{R}_2&\bar{R}_1
\end{array}\right)
&\left (\begin{array}{ll} \fe \\ \cfe
\end{array}\right)
\end{eqnarray}
for some constants $R_1, R_2\in\C$
due to the anticommutativity of the fermions.
This is not the only solution
to the exact equations of motion.
In general, the equation of motion
can be nonlinear, including terms like
$\cfe\dot{\fe}\dot{\cfe}$ etc.,
but these terms will not appear in
the iteration procedure outlined in previous sections.
Analogous to (\ref{hfn}), for higher derivatives we have
\be
\left (\begin{array}{ll} \fe^{(n)} \\ \cfe^{(n)}
\end{array}\right)
 = \left (\begin{array}{ll} R_1^{(n)}&R_2^{(n)} \\
\bar{R}_2^{(n)}&\bar{R}_1^{(n)}
\end{array}\right)
\left (\begin{array}{ll} \fe \\ \cfe
\end{array}\right) , \label{RR}
\ee
where
\begin{eqnarray}
\left (\begin{array}{ll} R_1^{(n)}& R_2^{(n)}\\ \bar{R}_2^{(n)}
&\bar{R}_1^{(n)}
\end{array}\right)
& \equiv \left (\begin{array}{ll} R_1&R_2 \\ \bar{R}_2&\bar{R}_1
\end{array}\right)^n.
\end{eqnarray}

%\be
%\left\{\begin{array}{ll} \fe^{(n)}=(-im-i\lambda R)^n\fe=(-im')^n
%\fe
%\\
%\cfe^{(n)}=(im+i\lambda R)^n\cfe=(im')^n \cfe
%\end{array}\right. .
%\ee
%Replacing it into the equation of motion, we find that $R$
%satisfied the following equation
%\be
%R=\sum_{j=0}(-i)^kC_{jk}[i(m+\lambda R)]^{j+k} . \ee
%We find the symplectic two-form as
Note that the effective conjugate momenta (\ref{pa})
must be linear in $\fe$ and $\cfe$
\be
q_{\fe} = i\left( \lam b_2 \fe +
\left( \frac{1}{2} + \lam b_1 \right) \cfe \right), \quad
q_{\cfe} = i\left( \left( \frac{1}{2} + \lam \bar{b}_1 \right) \fe
+ \lam \bar{b}_2 \cfe \right),
\ee
because there is no nonvanishing cubic term in $\fe$ and $\cfe$.
We have imposed the relation $\bar{q}_{\fe} = -q_{\cfe}$ because
$q_{\fe}\d\fe + q_{\cfe}\d\cfe$ should be Hermitian.
The symplectic 2-form on the reduced phase space
is thus of the form
\begin{eqnarray}
\label{Om2}
\Om=
i\left( (1+\lambda (b_1+\bar{b}_1))d\cfe d\fe
+\lambda b_2 d\fe d\fe + \lambda\bar{b}_2 d\cfe d\cfe \right),
\end{eqnarray}
for some constants $b_1, b_2\in\C$.
%where \bea  b_1=
%\sum_{i=0}^{\infty}\sum_{n=0}^{\infty}\sum_{j=i+1}^{\infty}
%(-1)^{j-i-1}\left[
%\left(C^{(1)}_{nj}\bar{R}_1^{(j-i+n-1)}+
%2C_{nj}^{(2)}R_2^{(j-i+n-1)}\right)R_1^{(i)}\right.\nonumber\\-
%\left.\left(C_{jn}^{(1)}R_2^{(j-i+n-1)}+
%2C_{jn}^{(3)}\bar{R}_2^{j-i+n-1}\right)\bar{R}_1^{(i)}\right],
%\eea \bea b_2=
%\sum_{i=0}^{\infty}\sum_{n=0}^{\infty}\sum_{j=i+1}^{\infty}
%(-1)^{j-i-1}\left[\left(C^{(1)}_{nj}R_2^{(j-i+n_1)}+
%2C_{nj}^{(2)}R_1^{(j-i+n-1)}\right)R_1^{(i)}\right.\nonumber\\-
%\left.\left(C_{jn}^{(1)}R_1^{(j-i+n-1)}+
%2C_{jn}^{(3)}R_2^{j-i+n-1}\right)\bar{R}_2^{(i)}\right],
%\eea
%It follows that the Poisson brackets should be
%\be
%\label{qqdot2}
%(\cfe, \fe)= -ib^{-1}(1-2\lambda b_1),
%\quad(\fe,\fe)=2\lambda
%b^{-1}b_2^{\ast},\quad(\cfe,\cfe)=-2\lambda b^{-1} b_2,
%\ee
%where $b\equiv det(2\Omega) = ...$.

By a change of variables
\be \label{cv}
\psi=\gamma_1\fe+\gamma_2\cfe
%-i\lambda(b_1^{\ast}\fe+b_2^{\ast})\cfe
,\quad
\bar{\psi}=\bar{\gamma}_2\fe+\bar{\gamma}_1\cfe ,
%\cfe+i\lambda(b_1\cfe+b_2\fe)
\ee
where
$\g_1$, $\g_2$ satisfy
\footnote{
A solution to (\ref{gg}) exists
if $2|\lam b_2|<1+\lam(b_1+\bar{b}_1)$.
This holds when $\lam$ is sufficiently small.
Otherwise the perturbative approach breaks down.
}
\be \label{gg}
|\g_1|^2 + |\g_2|^2 = 1+\lam(b_1+\bar{b}_1), \quad
\g_1\bar{\g}_2 = \lam b_2,
\ee
we have $\Om = id\bar{\psi} d\psi$ and so the Poisson bracket is standard
\be
\label{qqdot3} (\bar{\psi}, \psi)= -i,
\ee
and others are zeros.

To derive the Hamiltonian, we note that
by substituting all derivatives of the fermions
according to (\ref{RR}),
the potential becomes
\be
[V] = c \bar{\psi}\psi
\ee
for some real constant $c$
up to a constant.
The effective Hamiltonian (\ref{eH})
is also a function of $\psi$, $\bar{\psi}$,
which are the only two variables in the reduced phase space.
Hence
\be
H = m' \bar{\psi} \psi ,
\ee
where the effective mass $m'$ is given by
\be
m' = (m+\lam(c+i(b_1 R_1-\bar{b}_1\bar{R}_1-b_2 R_2+\bar{b}_2\bar{R}_2)))
(|\g_1|^2 - |\g_2|^2)^{-1}.
\ee
The effective Lagrangian is thus
\be \label{LL2}
\tilde{L} = i\bar{\psi}\dot{\psi} - m'\bar{\psi}\psi,
\ee
which is simplified by integration by parts.

Notice that the change of variables (\ref{cv}),
similar to a Bogoliubov transformation,
results in a change of vacuum upon quantization.
If the coefficients of higher derivative terms
depend on some background fields,
variation of the background fields will induce
the creation of particles.
The higher derivative terms also contribute to
the effective mass $m'$.

In the above we see that for two real fermions
only the quadratic terms in the potential have
some effect in the perturbative approach.
In general, if the number of independent fermion
degrees of freedom is finite, say $N$,
in the perturbative potential we can ignore
all interaction terms with more than $N$ factors of fermions
since they will all vanish on the reduced phase space.
Hence it is possible that a lot of information is lost in
the perturbative expansion.

\subsection{Supersymmetric Spacetime Noncommutative Field Theory}
\label{STNC}

Noncommutative field theories \cite{DN}
have attracted much attention in recent years
because of its natural appearance in string theory
as the low energy description of D-branes
in a $B$ field background \cite{NC}.
Compared with spatial noncommutativity,
field theories with spacetime noncommutativity
is much less understood \cite{NCST},
but also particularly interesting in the context of string theory \cite{SST}.
In terms of the Moyal product,
spacetime noncommutativity means infinite time derivatives.
In \cite{CHY}, perturbative approach is applied to
the spacetime noncommutative field theory of a scalar field.
It would be interesting to consider
noncommutative field theories with supersymmetry \cite{CR,CZ}
since supersymmetry tends to cure the UV/IR connection problem
in the quantum theory \cite{UVIR}.
As an example,
consider the Wess-Zumino model on
$1+1$ dimensional noncommutative spacetime
for a real scalar field $\Phi$ and Majorana fermion $\Psi$
with the Lagrangian density
\begin{eqnarray}
L&=&-\frac{1}{2}\partial_\mu\Phi\ast\partial^\mu\Phi
-\frac{i}{2}\bar{\Psi}\ast\g^\mu\partial_\mu\Psi
-\frac{1}{2}( m\Phi+\lambda\Phi*\Phi )^{*2}
\nn \\
&&-\frac{i}{2}m\bar{\Psi}*\Psi
-\left(\frac{i}{2}\lambda\bar{\Psi}*\Psi*\Phi+h.c.\right), \label{LWZ}
\end{eqnarray}
where $\g^{\mu}=(i\sigma^2, \sigma^1)$
and $\sigma^i$ are the Pauli matrices.

The $\ast$-product is defined by
\be
\label{star}
f\ast g(x)=
e^{\frac{i}{2}\th^{\mu\nu}\del_{\mu}\del'_{\nu}}
f(x)g(x')|_{x'=x}
\ee
and so
\be
\label{tx} [t,x]_{\ast}=i\th.
\ee

Up to total derivatives which do not change the action,
the Lagrangian (\ref{LWZ}) can be simplified as
\begin{eqnarray}
L&=&-\frac{1}{2}\partial_\mu\Phi\partial^\mu\Phi
-\frac{i}{2}\bar{\Psi}\g^\mu\partial_\mu\Psi
-( m\Phi+\lambda\Phi*\Phi )^2
\nn \\
&&-\frac{i}{2}m\bar{\Psi}\Psi
-i\lambda(\bar{\Psi}*\Psi)\Phi.
\label{LWZ1}
\end{eqnarray}

By a field redefinition analogous to (\ref{mom}),
to the lowest order
the effective Lagrangian is given by
\begin{eqnarray}
\widetilde{L}&\simeq& -\frac{1}{2}
\partial_\mu\varphi\partial^\mu\varphi
-\frac{i}{2}\bar{\psi}\g^\mu\partial_\mu\psi
-(m\varphi+\lambda\varphi\bar{*}\varphi)^2 \nn \\
&&-\frac{i}{2}m\bar{\psi}\psi
-i\lambda(\bar{\psi}\bar{*}\psi)\varphi,
\label{lsa}
\end{eqnarray}
where the $\bast$-product is defined as the $\ast$ in (\ref{star})
with the replacement
\bea
\del_t^n\varphi&\rightarrow&
\left\{\begin{array}{ll}
(\del_x^2-m^2)^{n/2}\varphi & (n=\mbox{even}), \nn \\
(\del_x^2-m^2)^{(n-1)/2}\del_t\varphi & (n=\mbox{odd}),
\end{array}\right.
\label{nbit}
\\
\del_t^n\psi&\rightarrow&
\left\{\begin{array}{ll}
(\del_x^2-m^2)^{n/2}\psi & (n=\mbox{even}), \nn \\
(\del_x^2-m^2)^{(n-1)/2}\g^0(\g^1\del_x + m)\psi & (n=\mbox{odd}),
\end{array}\right.
\label{delt}
\eea

%A problem with this example is that,
%had we chosen another way to rewrite (\ref{LWZ})
%by different integration by parts,
%the effective action at the first order approximation can be different.
%The perturbative approach only guarantees
%the equations of motion to be correct,
%but the effective action may not be unique.
%A related problem is that
%the $\bast$-product is not associative.

Note that $\ast$ is related to $\bast$
by the equations of motion at the lowest order.
This means that the supersymmery of (\ref{LWZ})
guarantees the supersymmetry of (\ref{lsa}) on shell
to the first order if we take the new fields $\varphi$, $\psi$
to transform in the same way as $\Phi$, $\Psi$.

As we mentioned in sec.\ref{Canonical},
the Hamilonian is always unbounded from below
in the canonical formulation,
and is thus in contradiction with the usual belief
that the Hamiltonian is positive definite
as a result of the superalgebra $H \sim Q^2$.
The only place that could go wrong in the usual argument
is the assumption of the supercharge $Q$ to be Hermitian.
Indeed, in an example in sec.\ref{Canonical}
we showed that quantization of the fermions destroys reality conditions.
Here we see that the perturbative approach not only
provides a consistent quantization but also preserves
supersymmetry order by order perturbatively.

\subsection{String Field Theory}

As we mentioned in the introduction,
all fields in the interaction term
in Witten's bosonic open string field theory \cite{Witten}
are modified by
\be \label{f}
f \rightarrow \tilde{f} \equiv e^{a^2\del_{\mu}\del^{\mu}}f,
\ee
where $a^2 = \ln(3\sqrt{3}/4)\a'$.
Similarly, in the bosonic closed string field theory \cite{BCSFT},
the same form (\ref{f}) appears with
$a^2 = \frac{1}{2}\ln(3\sqrt{3}/4)\a'$ \cite{KS2}.

The exponent $\del\cdot\del \simeq E^2-p^2$
implies that, after Wick rotation,
contribution from the UV excitations of string theory is suppressed.
It helps string theory to avoid the UV divergences
present in most ordinary field theories.
Another way to look at the effect of (\ref{f})
is to view $\tilde{f}$ as the fundamental field variable,
and its propagator is modified
\be
\frac{1}{p_E^2+m^2} \rightarrow
\frac{1}{p_E^2+m^2}e^{-2a^2 p_E^2},
\ee
which also implies a suppression of high energy modes.
However, the appearance of the higher derivative terms also
means that we are not sure how to treat this theory exactly \cite{EW}.

For Witten's bosonic open string field theory,
the technique of level truncation was shown to
be very effective in the calculation of tachyon potential \cite{KS,TP}.
In this calculation only the zero mode of each field has
to be considered.
A possible, although only weakly supportive reason why
the level truncation technique is effective
is that the coupling constants
are suppressed by a factor of $4/3\sqrt{3} \simeq 0.77$
when we increase the level number \cite{KS}.

According to the perturbative approach,
to the lowest order, we simply replace
the factor $\del_\mu\del^\mu$ in (\ref{f})
by the mass squared of the field $f$.
For instance, the level (0,0) truncation gives
the tachyon potential
\bea
V &=& -\frac{1}{2\a'}\Phi^2 +
\frac{1}{3}\left
(\frac{3\sqrt{3}}{4}\right)^3\tilde{\Phi}^3 \nn \\
&\rightarrow& \frac{1}{2\a'}\varphi^2 + \frac{1}{3}\varphi^3,
\eea
where we have used our results of first order approximation
in the perturbative approach to replace $\Phi$ by $\varphi$
according to (\ref{mom})
and $\del\cdot\del$ by $m^2 = -1/\a'$
for the tachyon field $\Phi$.
Note that this replacement is not suitable for
the zero mode of $\Phi$ because the zero mode
is determined solely by the potential term,
and is thus a nonperturbative effect.
In principle, the zero mode should be treated first
before the perturbative approach is applied.
Here we are considering fluctuations around the false vacuum.

For a field with $m^2 = n/\a'$,
from (\ref{f}) we have
\be \label{F}
e^{a^2\del_\mu\del^\mu} \rightarrow
\left(\frac{3\sqrt{3}}{4}\right)^n.
\ee
This implies that the interaction
for higher excitation modes are strengthened by
factors of $3\sqrt{3}/4$ when we increase the level number.
Therefore, combined with the decrease in coupling,
fluctuation modes other than the zero modes
will have roughly the same effective interaction
when we increase the level number.
If the reason why level truncation works for tachyon potential
is really the one mentioned above,
we will not expect level truncation to be effective
for fluctuation modes.

\section{Extension of the Perturbative Approach} \label{Extension:}

In this section we generalize our formal proof of all orders
in sec. \ref{AllOrder} to the situation where
the reduced phase space is allowed to keep
derivatives of fields up to an arbitrary given order.
For each field (boson or fermion)
$q_a$ we specify an integer $K_a > 1$
so that $q_a, \dot{q}_a, \cdots, q^{(K_a-1)}_a$
are kept in the reduced phase space.
Using the equations of motion,
in principle we can rewrite the $K_a$-th derivative of $q_a$
as a function on the reduced phase space
\be
q^{(K_a)}_a=f_{a}(q^{(j)}_b) , \label{ncb6}
\ee
where $j$ is less than $K_b$,
analogous to (\ref{qf}).
%The relation can be derived by varies methods
%different from the perturbation.

Higher derivatives of $q_a$ can be derived from (\ref{ncb6})
\be
q^{(i)}_a=f_{ai}(q^{(j)}_b),
\ee
and the functions $\f_{ai}$ can
be obtained recursively
\be
\f_{a(i+1)}=\left[\frac{d}{dt}\f_{ai}\right]
=\sum_b\sum_{j=0}^{k_b-2}
\frac{\del \f_{ai}}{\del q_b^{(j)}}q_b^{(j+1)}
+\sum_b\frac{\del \f_{ai}}{\del q_b^{(K_b-1)}}f_b ,
\ee
where we used the notation
\be
[A]\equiv A|_{q_a^{(i)}=f_{ai}}.
\ee

%Here are a few identities that will come in handy in the
%following.
From (\ref{bm}), (\ref{fm}), we find
\be
\label{ncb1}
\dot{p}_{ai}=\frac{\del L_0}{\del q_{a}^{(i)}}-p_{a(i-1)},
\ee
where $p_{ai}$ is the ith canonical momentum of $q_a$.
\\
For an arbitrary function $A$ on the phase space,
we find the following identities
\bea
\frac{d}{dt}[A]&=&\left[\dot{A}\right]+
\sum_{a}\frac{\del[A]}{\del q^{(K_a-1)}_a}(q^{(K_a)}_a-f_a),
\label{ncb2}
\\
\frac{\del[A]}{\del q^{(i)}_a}&=&
\sum_{b=1}\sum_{j=0}\left[\frac{\del A}{\del
q^{(j)}_b}\right]\frac{\del f_{bj}}{\del q^{(i)}_a}.
\label{ncb3}
\eea

The perturbative momentum $\Pi_{ai}$ can be read off from
\be
\sum_{b,j}p_{bj}\d q_b^{(j)}= \sum_{a}\sum_{i=0}^{K_a-1}\Pi_{ai}\d
q_a^{(i)}, \ee and we find
\be
\Pi_{ai}=\sum_{b,j}\left[
p_{bj}\frac{\partial q^{(j)}_b}{\partial q_a^{(i)}}\right] .
\label{ncb4}
\ee
The Hamiltonian is
\be
H=\left[\sum_{a}\sum_{i=0}^{K_a-1}\Pi_{ai}q_a^{(i+1)}-L\right].
\ee
 The symplectic structure is
  \bea
   \Omega &=& \sum_{b,j}dp_{bj}
  dq^{(j)}_b \nn \\ &=& \sum_a \sum_{i=0}^{K_a-1} d \Pi_{ai}d
  q_a^{(i)} \nn \\ &=& \sum_a\sum_{i=0}^{K_a-1}
  \sum_b\sum_{j=0}^{K_b-1}dq^{(i)}_a \frac{(-1)^a}{2}\left(
  (-1)^{a\cdot b}\frac{\del\Pi_{bj}}{\del q_a^{(i)}}
  -\frac{\del\Pi_{ai}}{\del q_b^{(j)}} \right) dq^{(j)}_b \nn \\
  &=&\sum_a\sum_{i=0}^{K_a-1}
  \sum_b\sum_{j=0}^{K_b-1}dq^{(i)}_a\Omega_{a_ib_j}dq^{(j)}_b , \eea
where
\be
\Omega_{a_ib_j}=\frac{(-1)^a}{2}\left((-1)^{a\cdot b}
\frac{\del\Pi_{bj}}{\del q_a^{(i)}} -\frac{\del\Pi_{ai}}{\del
q_b^{(j)}}\right).
\label{sympall}
\ee
In the above, when $a$, $b$ appear as the power of $(-1)$,
they are identified with $0$ or $1$ depending on
whether $q_a$ is a boson or fermion.

The Hamilton equations based on the symplectic structure
(\ref{sympall}) are
\be
(\Pi_{ai})^.=\sum_{b}\left\{ -\left(\Pi_{b(K_b-1)}\frac{\del
f_b}{\del q^{(i)}_a} \right) +(-1)^{(a\cdot b)}
\frac{\Pi_{b(K_b-1)}}{\del q^{(i)}_a}(q^{(K_b)}_b-f_b)\right\}
+\frac{\del [L]}{\del q^{(i)}_a}-\Pi_{a(i-1)}. \label{ncb5}
 \ee
  With
the help of (\ref{ncb1})-(\ref{ncb3}), one can show from
(\ref{ncb4}) that (\ref{ncb5}) is equivalent to the relation
(\ref{ncb6}) for $i \neq 0$ , and the case $i=0$ gives only an
identity.

\section*{Acknowledgement}

The authors thank Koji Hashimoto, Marc Henneaux and John Wang
for helpful discussions. This work
is supported in part by the National Science Council, the CosPA
project of the Ministry of Education, the National Center for
Theoretical Sciences, Taiwan, R.O.C. and the Center for
Theoretical Physics at National Taiwan University.

\end{document}